\documentclass{article}

\usepackage{fullpage}
\usepackage{natbib}

\title{Revenue Non-monotonicity in Matching Markets}
\author{Jason Hartline}
\date{}

\begin{document}

\maketitle

\begin{abstract}
  The Vickrey-Clarke-Groves (VCG) mechanism is infamously revenue
  non-monotone in combinatorial auctions.  I.e., when a buyer increases
  their value for a bundle of items, the total auction revenue may decrease.
  Combinatorial auctions exhibit complementarities which broadly
  result in complexities in auction theory.  This brief note shows
  that non-monotonicity in multi-item auctions is not a result of
  complementarities, and in fact, VCG is revenue non-monotone even in matching
  markets.
\end{abstract}

In auctioning multiple items to multiple buyers, the
Vickrey-Clarke-Groves (VCG) mechanism maximizes welfare in dominant
strategies.  It elicits values from the buyers; it selects the outcome
that maximizes the sum of the values of the buyers; and it charges
buyers their externalities.  An important application of the VCG
mechanism is in combinatorial auctions, where buyers have preferences
over bundles of items, i.e., when the items can be {\em complements}.

Combinatorial auctions introduce complexity into
auction theory, due to the possibility of complements in the
preferences of the buyers.  Even in the simplest case of single-minded
combinatorial auctions---where each buyer desires all of a single
specific subset of items and only that subset---\citet{AM-06}
infamously exhibit one of the complexities as a {\em revenue
non-monotonicity} \citep[see also,][]{RCL-07}.  Revenue non-monotonicity
is a situation where increasing a buyer's value for a bundle of items
results in the total revenue of the auction decreasing.

The example in \citet{AM-06} is:
\begin{itemize}
\item Buyer A values item 1 alone at \$1,
\item Buyer B values item 2 alone at \$0, and
\item Buyer C values the bundle $\{1,2\}$ at \$1.
\end{itemize}
The VCG outcome is to allocate the desired bundle to either A or C
and charge the winner \$1. The VCG revenue is \$1. Either buyer can be served, and the
externality is \$1 for not serving the other buyer.  Now consider
increasing Buyer B's value:
\begin{itemize}
\item Buyer B values item 2 alone at \$1.
\end{itemize}
Now the VCG outcome is to allocate the desired bundles to both A and B
(which are simultaneously compatible) and to charge each zero.  The VCG revenue is \$0.  To derive these payments, notice
that B's externality is zero.  When we serve B in this allocation, the
remaining buyers have a total surplus of \$1 (by serving A). When we
do not serve B, the highest surplus we can get from A and C is \$1
since they cannot be served together.  The difference in these two
quantities is B's externality and it is zero.  Hence, the VCG payment
of B is zero.  By symmetry A's payment is also zero.

In this brief note, I will show that the revenue non-monotonicity of
VCG is not a consequence of complements in the buyers' preferences for
items.  In fact, even when the items are {\em substitutes}, VCG can be
revenue non-monotone.  Because complements in preferences make
auctions that accommodate such preferences complex, understanding the
absence of complements is important.  For example, a buyer has {\em
  gross substitute} preferences if an increase in the price of one item does not
reduce the demand for another item \citep{GS-99}.  Unit-demand
preferences, where buyers have values for each item and their value
for a bundle of items is the maximum value of any item in the bundle,
are a special case of substitutes.

A unit-demand unit-supply matching market is given by $n$ buyers and
$m$ items where buyer $i$ has value $v_{ij} \geq 0$ for item $j$.  Feasible
outcomes are matchings where each buyer is matched to at most one item
and each item is matched to at most one buyer.  Here is a simple
example showing that VCG is revenue non-monotone in matching markets:

\begin{itemize}
\item Buyer A's values are $(\$2,\$0)$ for items $(1,2)$, and
\item Buyer B's values are $(\$1,\$0)$ for items $(1,2)$.
\end{itemize}
When both buyers only value item 1, the VCG outcome is the same as the
second-price auction for that item.  The VCG revenue is \$1.
Now consider increasing buyer $B$'s value:
\begin{itemize}
\item Buyer B's values are $(\$1,\$2)$ for items $(1,2)$.
\end{itemize}
Now the VCG revenue is 0. Both buyers can get their preferred outcome
without competing with the other buyer, i.e., each buyer's externality is
zero.

\citet{GS-99} characterized the VCG prices in matching markets as the
      {\em minimum Walrasian prices}.  Walrasian prices are ones where
      (a) everyone can be assigned one of their favorite items
      (a maximizer of value minus price) and (b) unsold items have
      price 0. Thus, we can reinterpret the above example as showing
      that the minimum Walrasian prices are revenue non-monotone.

      Viewed this way, it might seem as if this non-monotonicity is a
      result of a maximization problem within a minimization problem.
      This interpretation is not correct.  The maximum Walrasian
      prices are also revenue non-monotone.  Here is a simple example
      showing that the maximum Walrasian prices can be revenue
      non-monotone in matching markets:

      \begin{itemize}
      \item Buyer A's values are $(\$0,\$0)$ for items $(1,2)$, and
      \item Buyer B's values are $(\$1,\$0)$ for items $(1,2)$.
      \end{itemize}
      In this example, clearly the maximum Walrasian prices are $(\$1,\$0)$ and have a revenue of \$1.  Now consider increasing Buyer B's value:
\begin{itemize}
\item Buyer B's values are $(\$1,\$1)$ for items $(1,2)$.
\end{itemize}
Now the maximum Walrasian prices are $(\$0,\$0)$ for a total revenue of \$0.
This follows because asymmetric prices result in both buyers desiring
the lower price, so the prices must be equal.  For Buyer A to want any item, the price must be $0$.

The second example of revenue non-monotonicity of maximum Walrasian
prices is somewhat related to a non-monotonicity result of
\citet{HR-15}.  They show that the expected revenue from the
multi-item single-buyer mechanism that maximizes its expected revenue
when the buyer's values are drawn from a known distribution is revenue
non-monotone in the distribution.  This problem is equivalent to the
problem of maximum Walrasian prices without the constraint that unsold
items have price zero.  However, note that in the second example
above, if we relax the constraint that unsold items have price zero
and imagine a single buyer that is either type A or type B with equal
probability, then there is no longer a revenue non-monotonicity.
Thus, the revenue non-monotonicity of maximum Walrasian prices is a
different phenomenon.

In conclusion, revenue non-monotonicity is not a complexity that
arises from complements in preferences; it can also occur even in
multi-item auctions with substitutes, such as the canonical matching
market.

\bibliographystyle{apalike}
\bibliography{refs}

\begin{thebibliography}{}

\bibitem[Ausubel and Milgrom, 2006]{AM-06}
Ausubel, L.~M. and Milgrom, P. (2006).
\newblock The lovely but lonely vickrey auction.
\newblock In Cramton, P., Shoham, Y., and Steinberg, R., editors, {\em
  Combinatorial Auctions}, pages 17--40. MIT Press.

\bibitem[Gul and Stacchetti, 1999]{GS-99}
Gul, F. and Stacchetti, E. (1999).
\newblock Walrasian equilibrium with gross substitutes.
\newblock {\em Journal of Economic theory}, 87(1):95--124.

\bibitem[Hart and Reny, 2015]{HR-15}
Hart, S. and Reny, P.~J. (2015).
\newblock Maximizing revenue with multiple goods: Nonmonotonicity and
  submodularity.
\newblock {\em Journal of Economic Theory}, 157:116--132.

\bibitem[Rastegari et~al., 2007]{RCL-07}
Rastegari, B., Condon, A., and Leyton-Brown, K. (2007).
\newblock Revenue monotonicity in combinatorial auctions.
\newblock In {\em Proceedings of the 22nd National Conference on Artificial
  Intelligence}, volume~1, pages 122--127.

\end{thebibliography}

\end{document}